\documentclass[%
 reprint,
 amsmath,amssymb,
prb,
]{revtex4-1}
\usepackage{graphicx}
\usepackage{dcolumn}
\usepackage{bm}
\usepackage{braket}
\usepackage{color}
\usepackage{accents}
\usepackage{enumitem}

\begin{document}
\title{
Anomalous proximity effect of a spin-singlet superconductor\\
with a spin-orbit interaction
}
\author{Jaechul Lee$^{1}$}
\author{Satoshi Ikegaya$^{2}$}
\author{Yasuhiro Asano$^{1}$}%
\affiliation{
$^{1}$Department of Applied Physics, Hokkaido University, Sapporo 060-8628, Japan\\
$^{2}$ Institute for Advanced Research, Nagoya University, Nagoya 464-8601, Japan\\
}
\date{\today}
\begin{abstract}
The anomalous proximity effect of a spin-triplet $p$-wave superconductor has been known as 
 a part of the Majorana physics and is explained
by the penetration of zero-energy states from a surface of a superconductor to a dirty normal
metal.
We demonstrate that a spin-singlet $d$-wave superconductor without 
 any surface zero-energy states exhibits the anomalous 
proximity effect in the presence of a specific spin-orbit interaction.
The results show the quantization of the zero-bias conductance in a 
dirty normal-metal/superconductor junction.
We also discuss a relation between our findings and results in an experiment on a 
CoSi$_2$/TiSi$_2$ junction.
\end{abstract}
\maketitle

\section{Introduction}
When a superconductor (SC) is attached to a normal metal, Cooper pairs penetrate from the SC into the
 normal metal and modify its electromagnetic and thermal properties.
This phenomenon, known as the proximity effect, exhibits distinct behavior depending on the 
symmetry of the pair potential.
Specifically, the proximity effect of a spin-triplet $p$-wave SC indicates remarkable transport phenomena
such as 
the quantization of zero-bias conductance in a dirty normal-metal / superconductor (DN/SC) 
junction~\cite{tanaka:prb2004,tanaka:prb2005-2}
and the fractional current-phase relationship of a Josephson currents in a SC/DN/SC 
junction~\cite{asano:prl2006, asano:prb2006-2}.
These unusual phenomena are referred to as anomalous proximity effect (APE).

The APE is a result of the interplay between two interference effects: 
the proximity effect in a DN attached to a SC 
and the formation of Andreev bound states at the surface of a SC.
The presence or absence of the proximity in a DN depends sensitively on the 
symmetry of the pair potential~\cite{asano:prb2001d}.
To host Andreev bound states at the surface of a SC,
the pair potential is necessary to change its sign on the Fermi 
surface~\cite{buchholtz:prb1981, hu:prl1994, tanaka:prl1995, asano:prb2004}.
Symmetry analysis in the early stages of the study suggested that the
 APE is a phenomenon unique to spin-triplet SCs.~\cite{asano:prb2006-2} 
In addition to the conductance quantization in a DN/SC junction, the APE 
causes the zero-bias anomaly of the conductance spectra in a T-shaped junction~\cite{asano:prl2007tshape}, and 
the unusual surface impedance~\cite{asano:prl2011}.
Unfortunately, it would be very difficult to observe the APEs in experiments 
because spin-triplet SCs are very rare. 
A topological material based compound
Cu$_x$Bi$_2$Se$_3$ and
several uranium compounds such as 
UPt$_3$, UBe$_{13}$, UGe$_2$, and UTe$_2$
are candidates of the spin-triplet SC~\cite{stewart_84, Ott_83, saxena_00, ran_19, jiao_20, hor:prl2010, sasaki_11}. 
However, spin-triplet superconductivity in these materials are still under debate.

The fabrication of artificial spin-triplet SCs is an important issue 
these days to realize the quantum computation 
by applying non-Abelian statistics of 
Majorana Fermions.~\cite{sato:prb2006, sato:prb2009, lutchyn:prl2010, oreg:prl2010, asano:prb2013, Sarma_15}.
The APE is a part of the Majorana physics because Majorana zero modes are a special case of 
the Andreev bound states at the surface of a spin-triplet SC~\cite{asano:prb2013}.
 These theoretical studies have suggested 
 that spin-orbit interactions (SOIs) enable the realization of spin-triplet superconductivity 
 in a spin-singlet SC. 
Moreover, a theory shows that a nonzero integer number $\mathcal{N}_{\mathrm{ZES}}$, mathematically known as an Atiyah-Singer index, represents exactly the quantized value of the zero-bias conductance 
in a DN/SC junction~\cite{ikegaya:prb2015, ikegaya:prb2016}.
According to their argument, $\mathcal{N}_{\mathrm{ZES}}$ represents 
the number of zero-energy states that penetrate from a surface of SC into a DN and form the resonant transmission channels.
This conclusion leads us to infer that spin-triplet superconductivity is only a sufficient condition for 
$\mathcal{N}_{\mathrm{ZES}} \neq 0$. 
Two of authors looked for necessary conditions for the BdG 
Hamiltonian that provide a nonzero $\mathcal{N}_{\mathrm{ZES}}$~\cite{ikegaya:prb2018}.
We found that several Hamiltonians breaking time-reversal symmetry lead to a nonzero index
and they describe the artificial SCs hosting Majorana zero 
modes~\cite{alicea:prb2010, you:prb2013, ikegaya:prb2018}. 
In addition, we also found that a Hamiltonian for a spin-singlet $d_{xy}$-wave SC with a specialized SOI
gives a nonzero index.~\cite{lee-ikegaya:prb2021,ikegaya_21} 
It has been well established that a $d_{xy}$-wave SC 
without SOIs hosts highly degenerate zero-energy states at its clean surface 
parallel to the $y$ direction~\cite{buchholtz:prb1981, hu:prl1994, tanaka:prl1995, asano:prb2004}.
But in the absence of SOIs, $\mathcal{N}_{\mathrm{ZES}} = 0$ holds true, 
which means that zero-energy states are fragile under impurity scatterings.
A SOI transforms such fragile zero-energy states to robust zero-energy 
states.~\cite{lee-ikegaya:prb2021,ikegaya_21}

In 2021, an experiment observed a clear signal of APE~\cite{lin:sciadv2021}.
The conductance spectra in a T-shaped junction connecting to CoSi$_2$ grown on a Si substrate 
show the zero-bias anomaly which is a typical phenomenon of the APE.~\cite{asano:prl2007tshape}
Their transport measurement indicates strong SOIs near the interface
 between a thin CoSi$_2$ single-crystal and a Si substrate~\cite{chiu:cjp2024,chiu:nanoscale2023}.
However, spin-singlet $s$-wave superconductivity has been 
well-established in bulk CoSi$_2$~\cite{mattheiss_88, tsutsumi_95}.
Unlike a $d_{xy}$-wave SC, the zero-energy states are absent at the junction interface. 
At present, it is not clear if SOIs can cause the APE in a junction that consists
a spin-singlet SC without any surface Andreev bound states.
We discuss this issue in the present paper.

In this paper, we theoretically study the differential conductance in a DN/SC junction 
as shown in Fig.~\ref{fig:fig1}. 
We assume the spin-singlet $s$-wave and  spin-singlet $d_{x^2-y^2}$-wave pair potentials
in a SC. In the absence of SOIs, the Andreev bound states are absent 
at a junction interface. 
We introduce three types of SOIs in a SC and the nonmagnetic random impurity potential in a DN.
The conductance is calculated based on the Blonder-Tinkham-Klapwijk formula and 
the transport coefficients are obtained by using the recursive Green's function method.
A $d_{x^2-y^2}$-wave SC with a persistent spin-helix type SOI causes the APE. Unfortunately, 
an $s$-wave SC does not exhibit the APE with any types of SOIs.

This paper is organized as follows.
We explain our theoretical model in Sec.~\ref{sec:model}.
The results of the differential conductance are presented in Sec.~\ref{sec:results}.
We explain why a persistent spin-helix type SOI is necessary for the APE and why 
an $s$-wave SC does not indicate APE
in Sec.~\ref{sec:index}. The discussion of our results is presented in Sec.~\ref{sec:discussion}. 
The conclusion is given in Sec.~\ref{sec:conclusion}.

\section{Model}
\label{sec:model}
\begin{figure}[bbbb]
\begin{center}
    \includegraphics[width=0.8\columnwidth]{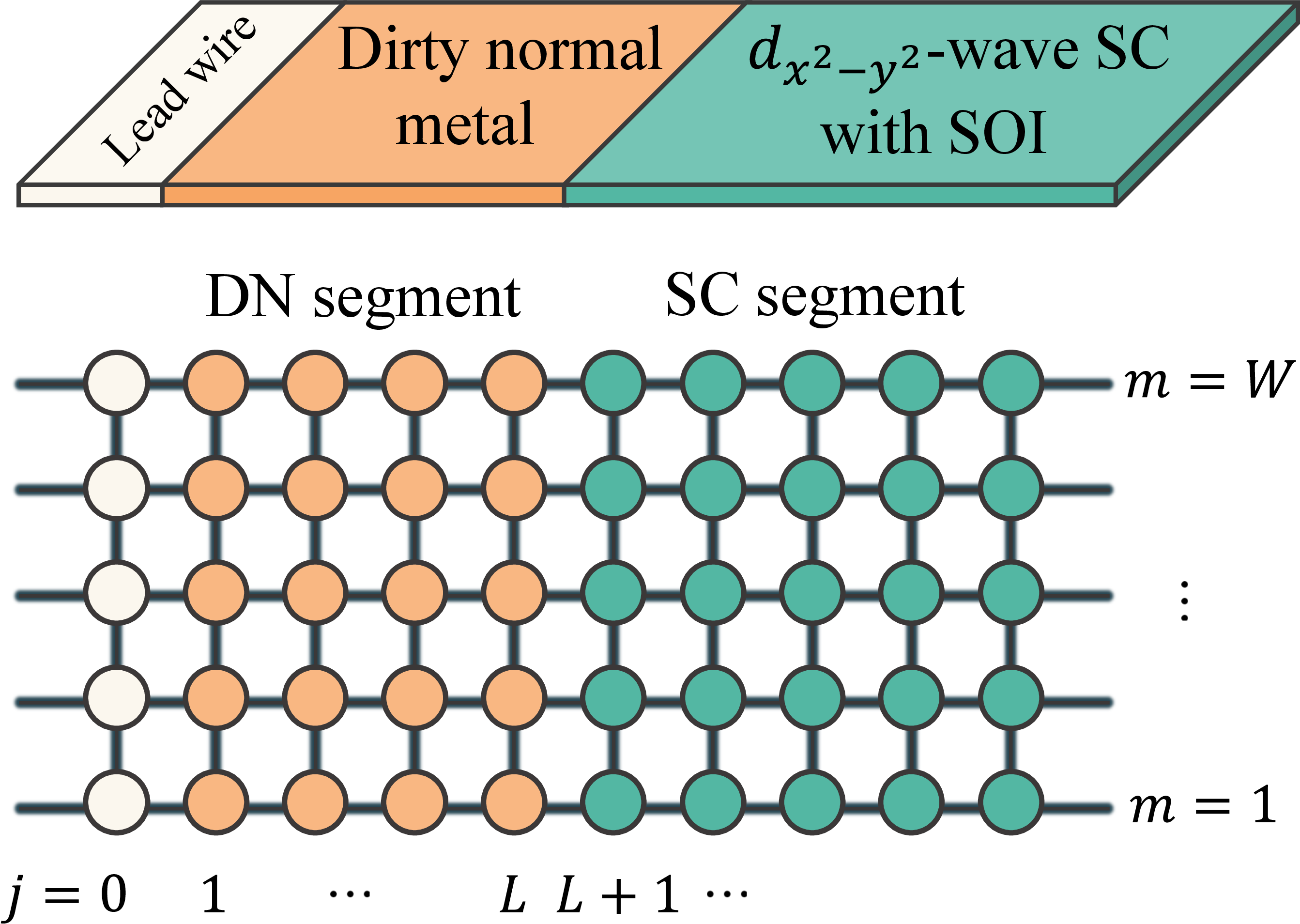}
    \caption{Schematic figure of a two-dimensional dirty normal-metal / SC junction.
We consider spin-singlet pair potentials ($s$-wave or $d_{x^2-y^2}$-wave) and 
three types of spin-orbit interactions in a SC.
We also introduce nonmagnetic random impurities in a normal metal.}
    \label{fig:fig1}
\end{center}
\end{figure}
We describe a DN/SC junction on a two-dimensional tight-binding lattice 
as shown in Fig.~\ref{fig:fig1}, 
where $L$ is the length of a DN, $W$ is the width of a junction, 
$\bm{x} (\bm{y})$ is the unit vector in the $x (y)$ direction, 
and a vector $\bm{r} = j \bm{x} + m \bm{y}$ indicates a lattice site.
The Hamiltonian consists of four terms,
\begin{align}
H = H_{\text{kin}} + H_{\text{imp}}  + H_{\text{SOI}} +  H_{\Delta}  
\end{align}
The kinetic energy of an electron is represented by
\begin{align}
H_{\text{kin}} = & - t \sum_{\bm{r}, \bm{r'}} \sum_{\alpha = \uparrow, \downarrow}
\left(
	c^{\dagger}_{\bm{r}, \alpha} c_{\bm{r'}, \alpha}
	+ c^{\dagger}_{\bm{r'}, \alpha} c_{\bm{r}, \alpha} 
\right)
\nonumber \\ 
& + (4t - \mu) \sum_{\bm{r}, \alpha} c^{\dagger}_{\bm{r}, \alpha} c_{\bm{r}, \alpha} ,
\end{align}
where $t$ is the nearest-neighbor hopping integral, $\mu$ is the chemical potential, and 
$ c^{\dagger}_{\bm{r}, \alpha} (c_{\bm{r},\alpha}) $ is
the creation (annihilation) operator of an electron with spin $\alpha$ at $\bm{r}$.
The second term represents the random impurity potential in a normal metal
\begin{align}
H_{\text{imp}} = \sum_{j = 1}^{L} \sum_{m = 1}^{W}  \sum_{\alpha}
V_{\bm{r}}  c^{\dagger}_{\bm{r},\alpha} c_{\bm{r}, \alpha},
\end{align}
where $V_{\bm{r}}$ is potential given randomly in the range of $-V_{\mathrm{imp}}/2 \leq 
V_{\bm{r}} \leq V_{\mathrm{imp}}/2$ by using random numbers with uniform distribution.
All the conductance shown below are obtained by averaging the conductance 
over a number of different samples. The ensemble average of the conductance is insensitive to 
 the types of the random numbers. Thus, we would reach the same conclusions 
 even if we used the random number with Gaussian distribution.
The superconducting segment $j \ge L+1$ is free from potential disorder.
We consider the SOI in a SC as
\begin{align}\label{eq:RSOI}
H_{\text{SOI}}= &\frac{i}{2} \sum_{\bm{r}, \alpha, \alpha^\prime}
\bigg[ 
	 \lambda_x \left(
		c^{\dagger}_{\bm{r}, \alpha} c_{\bm{r} + \bm{x}, \alpha^\prime}
		-c^{\dagger}_{\bm{r} + \bm{x}, \alpha} c_{\bm{r}, \alpha^\prime}
	\right) (\sigma_y)_{\alpha, \alpha^\prime}
\nonumber \\ 
	-  \lambda_y & \left(
		c^{\dagger}_{\bm{r}, \alpha} c_{\bm{r} + \bm{y}, \alpha^\prime}
		-c^{\dagger}_{\bm{r} + \bm{y}, \alpha} c_{\bm{r}, \alpha^\prime}
	\right) (\sigma_x)_{\alpha, \alpha^\prime}
\bigg]
, 
\end{align}
where $\lambda_{x(y)}$ represents the strength of SOI coupled to a momentum $k_x(k_y)$, 
and $\sigma_{i}$ for $i=x, y,$ and $z$ represents the Pauli matrix in spin space. 
In this paper, we 
mainly consider the three types SOI, 
\begin{subequations}
\label{eq:lambda}
\begin{eqnarray}
&(\lambda_x, \lambda_y) = (\lambda, 0) & \quad x\text{-type}, \label{eq:subeq1}\\
&(\lambda_x, \lambda_y) = (0, \lambda) & \quad y\text{-type}, \label{eq:subeq2}\\
&(\lambda_x, \lambda_y) = (\lambda, \lambda) & \quad \text{Rashba}. \label{eq:subeq3}
\end{eqnarray}
\end{subequations}
Here $x$-type and $y$-type SOI are a source of the persistent spin helix 
\cite{bernevig:prl2006,schliemann:rmp2017,kohda:semiscitech2017}.
The last one is the Rashba spin-orbit interaction.
The pair potential for a spin-singlet $d_{x^2-y^2}$ symmetry class is described by
\begin{align}
H_{\Delta} = 
&\frac{\Delta}{2} \sum^{\infty}_{j = L + 1} \sum^{W}_{m=1} 
\left(
c^{\dagger}_{\bm{r} + \bm{x}, \uparrow} c^{\dagger}_{\bm{r}, \downarrow} 
+ c^{\dagger}_{\bm{r}, \uparrow} c^{\dagger}_{\bm{r} + \bm{x}, \downarrow} \right. \\
&  - c^{\dagger}_{\bm{r} + \bm{y}, \uparrow} c^{\dagger}_{\bm{r}, \downarrow} 
 - c^{\dagger}_{\bm{r}, \uparrow} c^{\dagger}_{\bm{r} + \bm{y}, \downarrow} 
  \left. + \mathrm{H. c.} \right),
\end{align}
where $\Delta$ is the amplitude of the pair potential.
For a spin-singlet $s$-wave SC, we choose
\begin{align}
H_{\Delta} = 
&\Delta \sum^{\infty}_{j = L + 1} \sum^{W}_{m=1} 
\left[ c^{\dagger}_{\bm{r}, \uparrow} c^{\dagger}_{\bm{r}, \downarrow} + \mathrm{H. c.}\right].
\end{align}

The differential conductance of a DN/SC junction is calculated based on the 
Blonder-Tinkham-Klapwijk formula~\cite{blonder:prb1982}, 
\begin{align}
G_{\text{NS}} &(eV) = \frac{ e^2}{h} \sum_{l, l^\prime, \alpha, \alpha^\prime} \nonumber\\
\times &\left[ \delta_{l,l^\prime} \delta_{\alpha, \alpha^\prime} 
- |r^{\mathrm{ee}}_{l, \alpha; l^\prime, \alpha^\prime}|^2 
+ |r^{\mathrm{he}}_{l, \alpha; l^\prime, \alpha^\prime}|^2 \right]_{E=eV} ,
\end{align}
where $r^{\mathrm{ee}}_{l, \alpha; l^\prime, \alpha^\prime}$ is the normal reflection coefficient 
from the $l^\prime$ th propagating channel with spin $\alpha^\prime$ in the electron branch to 
 the $l$ th propagating channel with spin $\alpha$ in the electron branch, whereas
 $r^{\mathrm{he}}_{l,\alpha; l^\prime, \alpha^\prime}$ is the Andreev reflection coefficient 
from the $l^\prime$ th propagating channel with spin $\alpha^\prime$ in the electron branch to 
 the $l$ th propagating channel with spin $\alpha$ in the hole branch.
These reflection coefficients are calculated by using the recursive Green's function 
method~\cite{fisher_81, ando_91}.
The normal state conductance of a DN is calculated based on Landauer formula
\begin{align}
G_{\text{N}} & = \frac{ e^2}{h} \sum_{l, l^\prime, \alpha, \alpha^\prime} 
 |t_{l, \alpha; l^\prime, \alpha^\prime}|^2, \label{eq:gn}
\end{align}
where $t_{l, \alpha; l^\prime, \alpha^\prime}$ is the normal transmission coefficient
from the $l^\prime$ th propagating channel with spin $\alpha^\prime$ to 
 the $l$ th propagating channel with spin $\alpha$ through a DN.

In this paper, the energy is measured in units of $t$.  
We fix several parameters as $\mu = 2t$, $\Delta = 0.1t$, $L = 50$, and $W = 25$. 
The pair potential is not determined self-consistently because the gap equation always has 
a stable solution even in the presence of SOIs.
We use typically 100-500 different samples for the ensemble averaging over random impurity configurations.

\section{Results}
\label{sec:results}

\subsection{$d$-wave}
\label{ssec:d-wave}

\begin{figure}[tttt]
\begin{center}
    \includegraphics[width=0.45\textwidth]{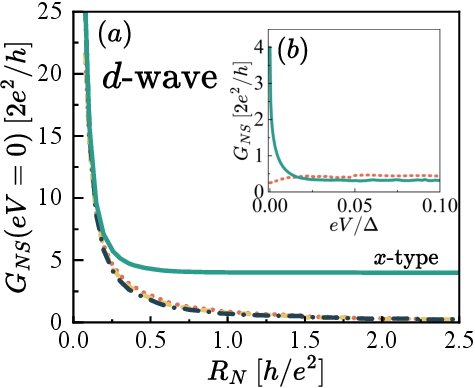}
    \caption{(a) The results for a $d_{x^2-y^2}$-wave junction.
	The zero-bias differential conductance is plotted
	as a function of the normal state resistance $R_{\mathrm{N}}$ in a normal metal. 
	The results for $x$-type SOI decreases to 4 $G_0$ at $R_{\mathrm{N}} \to \infty$.
	The results for $y$-type SOI, those for Rashba SOI, and those without SOI almost overlap 
	with one another. They decrease to zero at $R_{\mathrm{N}} \to \infty$.
	(b) The differential conductance is plotted as a function of the bias voltage
 at $R_N=4.5 (h/e^2)$.}
    \label{fig:fig2}
\end{center}
\end{figure}

We first study the conductance in a junction consisting of a $d_{x^2-y^2}$-wave SC.
In Fig.~\ref{fig:fig2}(a), the zero-bias conductance $G_{\text{NS}}(0)$ is plotted as a function of the
normal state resistance of a DN $R_{\mathrm{N}} = G^{-1}_{\mathrm{N}}$ for $\lambda = 0.5t$, where 
the vertical axis is normalized to $G_0 = 2e^2/h$.
The conductance is averaged over 100 samples with different random configuration.
The results are separated into two groups: $x$-type SOI and other cases.
The results for the $y$-type SOI, those for Rashba SOI, and those without SOI almost overlap with one another.  
They decrease with increasing $R_{\mathrm{N}}$ and vanish for large $R_{\mathrm{N}}$.  
The effects of the SOI on the zero-bias conductance are negligible for $y$-type SOI and Rashba SOI. 
These behaviors can be explained by the classical expression of the
total resistance of the resistors in series. 
Because the resistance in a SC is zero, the total resistance of the 
junction would be given by
\begin{align}
R_{\mathrm{NS}} = R_{\mathrm{B}} + \tilde{R}_{\mathrm{N}} = G^{-1}_{\mathrm{NS}}, \label{eq:classicalRns}
\end{align} 
where $R_{\mathrm{B}}$ is the normal resistance due to the potential barrier at the DN/SC interface.
In the present results the Sharvin resistance replaces $R_{\mathrm{B}}$ because we do not introduce
potential barrier at the interface. 
The usual proximity effect decreases the resistance in a DN to $\tilde{R}_{\mathrm{N}} 
\lesssim {R}_{\mathrm{N}}$ only slightly. As a result, 
The relation $G_{\mathrm{NS}} \to 0$ is expected in the limit of $ R_{\mathrm{N}} \to \infty$. 
On the other hand, the conductance for $x$-type SOI deviates from such a relationship 
and saturates at a finite value of $4 G_0$ for large $R_{\mathrm{N}}$.
Such unusual behavior is an aspect of the APE.~\cite{ikegaya:prb2016}.   
The resonant states at zero energy form the perfect transmission channels in a DN.
In Fig.~\ref{fig:fig2}(a), the number of such zero-energy states is 4. 
In Fig.~\ref{fig:fig2}(b), 
 the differential conductance $G_{\mathrm{NS}}$ at $R_{\mathrm{N}} = 4.5 (h/e^2)$ is plotted as a function of the 
 bias voltage $eV$. 
 The conductance for the $x$-type SOI decreases rapidly with increasing $eV$ because 
the perfect resonant transmission occurs only at zero bias. 
As a consequence, the results for the $x$-type SOI exhibit a sharp peak at zero-bias. 
For comparison, we plot the results for Rashba SOI in Fig.~\ref{fig:fig2}(b) with a broken line.
The conductance exhibits no distinct peak structures around zero bias.

\begin{figure}[tttt]
\begin{center}
    \includegraphics[width=0.34\textwidth]{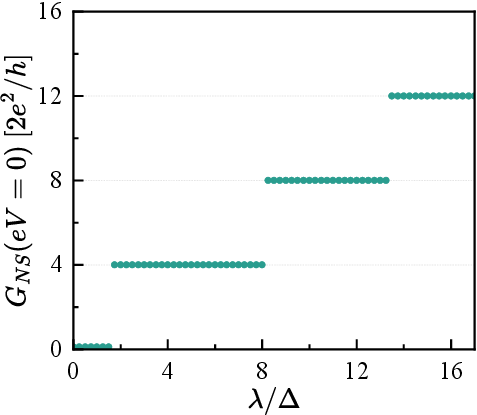}
    \caption{The zero-bias conductance of a $d_{x^2-y^2}$-wave junction for $x$-type SOI is plotted
 as a function of the strength of SOI $\lambda$ at $R_N= 4.5 G_0^{-1}$.}
    \label{fig:fig3}
\end{center}
\end{figure}

The zero-bias conductance in the limit of $R_{\mathrm{N}}\to \infty$ depends on 
the amplitude of the $x$-type SOI $\lambda$ as shown in
 Fig.~\ref{fig:fig3}, where $G_{\mathrm{NS}}(0)$ is plotted as a function of $\lambda$. 
The conductance remains zero for $\lambda < 0.175 t $ and jumps to a finite value 
of $4 G_0$ at $\lambda = 0.175 t$. Such step-like behavior is observed also at 
$\lambda = 0.85 t$ and $\lambda = 1.35 t$. 
The conductance is quantized at $ 4G_0$, 8 $G_0$, and 12 $G_0$, these 
steps. As we will discuss in Sec.~\ref{sec:index}, the minimum value of the conductance 
is given by $ G_0 \mathcal{N}_{\text{ZES}}$ where $\mathcal{N}_{\text{ZES}}$ 
is the number of zero-energy states that form the perfect transmission channels in a DN. 
The results indicate that $\mathcal{N}_{\text{ZES}}$ changes discontinuously by 4.

\subsection{$s$-wave}
\label{ssec:s-wave}

\begin{figure}[tttt]
\begin{center}
    \includegraphics[width=0.45\textwidth]{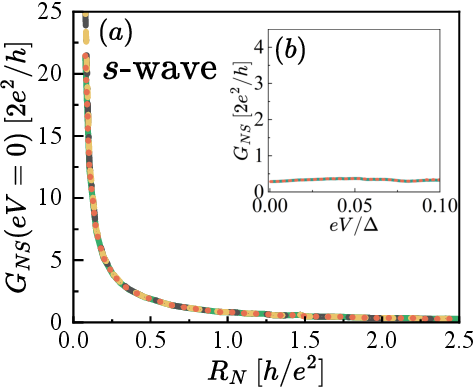}
    \caption{(a)
	The results for an $s$-wave DN/SC junction. 
   The zero-bias differential conductance is plotted 
	as a function of normal state resistance $R_{\mathrm{N}}$ in a normal metal. 
	Although we consider $x$- and $y$-types SOI, Rashba SOI, and absence of SOI, 
	all the results almost overlap with one another. 
	(b) The differential conductance is plotted as a function of the bias voltage
 at $R_N=4.5 (h/e^2)$. }
    \label{fig:fig4}
\end{center}
\end{figure}

Secondly, 
 we discuss the absence of the APE in a DN/SC junction for an $s$-wave symmetry. 
In Fig.~\ref{fig:fig4}(a), we plot the zero-bias conductance as a function of $R_{\mathrm{N}}$ 
for a $s$-wave superconductor including $x$-type SOI, $y$-type SOI, and Rashba SOI with $\lambda=0.5 t$.
We also plot the results without SOI $\lambda=0$ in the figure.
All of the results overlap with one another, which indicate that the effects of SOI 
on the conductance are negligible in an $s$-wave junction.
In all cases, the zero-bias conductance decreases to zero with increasing $R_{\mathrm{N}}$.
 In the inset of Fig.~\ref{fig:fig4}(b), we also plot the differential conductance 
$G_{\textrm{NS}}$ at $R_{\textrm{N}} = 4.5 \, G_0^{-1}$ as a function of the bias voltages.
The results show that the conductance is insensitive to the bias voltage.
The results in Fig.~\ref{fig:fig4} suggest that $s$-wave superconductor junctions 
do not indicate the APE regardless of the type of SOI.
We will discuss the reasons in Sec.~\ref{sec:index}.


\section{Modified pair potential and Index}
\label{sec:index}

There are two conditions for a superconductor that indicates the APE: 
the presence of the usual proximity effect in a DN and
the existence of the Andreev bound states at its surfaces parallel to the $y$ direction.~\cite{asano:prl2006}
We first consider a spin-singlet even-parity SC without SOIs. 
The BdG Hamiltonian is such a SC can be block diagonalized into two $2\times 2$ Hamiltonian.
When the pair potential in one spin sector is $\Delta(\bm{k})$, that in the other is $-\Delta(\bm{k})$.
A SC causes the usual proximity effect when its the pair potential 
satisfies~\cite{asano:prb2001d}
\begin{align}
\label{eq:proximity_effect}
\Delta (k_x , -k_y) \neq -\Delta (k_x, k_y), 
\end{align}
where $k_x$ and $k_y$ are the wave number on the Fermi surface.
The pair potentials considered in this paper are described by
\begin{align}
\Delta(\bm{k}) = \left\{ \begin{array}{ll}
 \Delta (k_x^2-k_y^2) / k_F^2  & :d_{x^2-y^2}\mathrm{-wave} \\
 \Delta & :s\mathrm{-wave}
 \end{array} \right., \label{eq:delta_k}
\end{align}
where $k_F$ is the Fermi wave number on the isotropic Fermi surface. 
Both $d_{x^2-y^2}$-wave pair potential and $s$-wave pair potential satisfy Eq.~\eqref{eq:proximity_effect}.
In the junction geometry in Fig.~\ref{fig:fig1}, the wave number in the $y$ direction 
$k_y$ indicates a transport channel.
Meanwhile, the presence of the surface Andreev bound states is ensured 
when the pair potential satisfies~\cite{tanaka:prl1995, asano:prb2004}
\begin{align} \label{eq:ABS}
\Delta (k_x , k_y) \Delta (-k_x, k_y) <0.
\end{align}
Either $d_{x^2-y^2}$-wave pair potential or $s$-wave pair potential do not satisfy 
the condition in the absence of SOIs.~\cite{asano:prl2006} 
Thus spin-singlet SCs do not indicate the APE in the absence of SOIs.

\begin{figure}[tttt]
\begin{center}
    \includegraphics[width=0.35\textwidth]{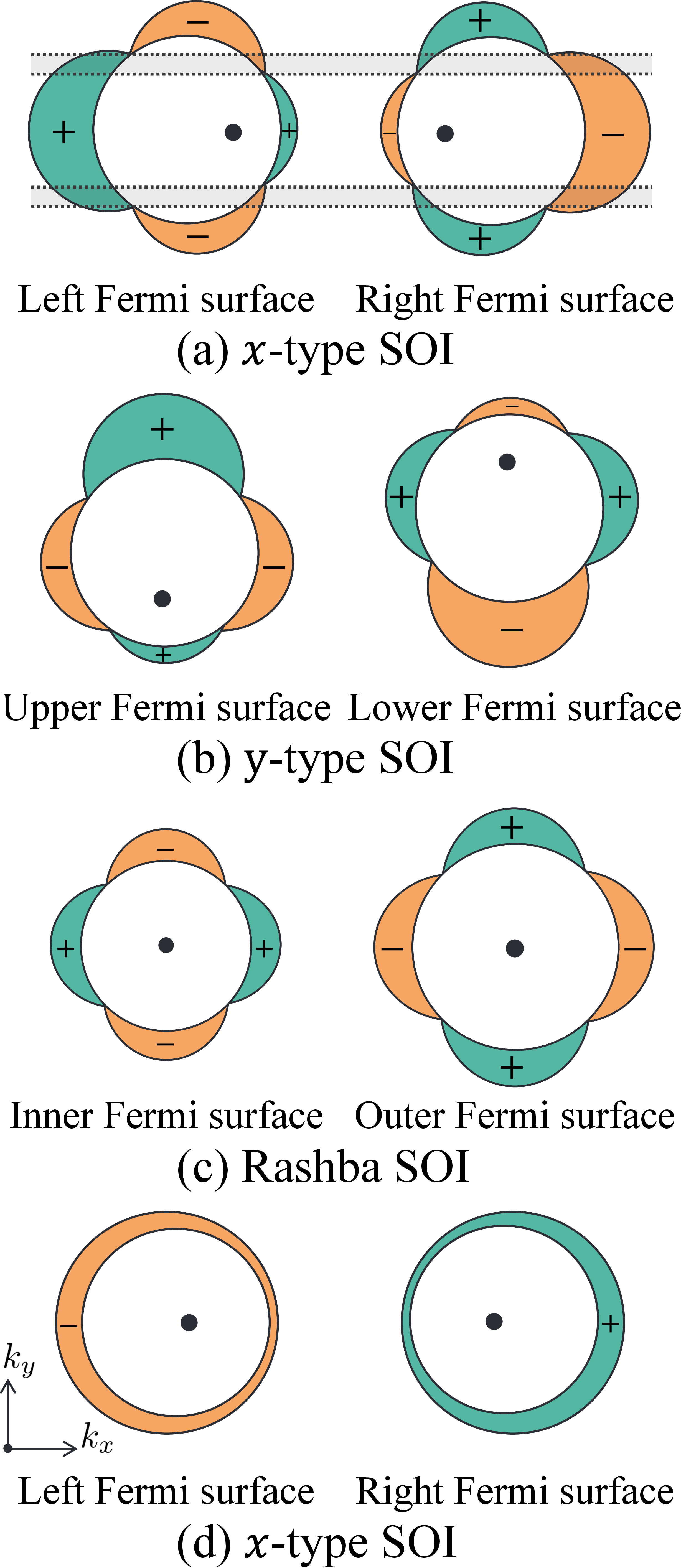} 
    \caption{ The pair potentials of a $d_{x^2-y^2}$-wave SC on the Fermi surface with (a) $x$-type SOI 
	and (b) $y$-type SOI, and (c) Rashba SOI. The black dot in the figure indicates the $\Gamma$ point ($\bm{k} = 0$).
	In the domains between the two dotted lines in (a), the condition $\Delta (k_x , k_y) \Delta (-k_x, k_y) <0$ is satisfied.
	In (d), the pair potentials of an $s$-wave SC are illustrated for $x$-type SOI.
	}
	\label{fig:fig5}
\end{center}
\end{figure}

Secondly, we discuss how SOI modifies 
the pair potential on the Fermi surface of a $d_{x^2-y^2}$-wave SC. 
The Hamiltonian considered in this paper is represented in continuous space
\begin{align}
H_{\text{BdG}}(\bm{k})&=( \xi_{\bm{k}} - \lambda_x k_x \hat{\sigma}_y) \hat{\tau}_z 
+\lambda_y k_y \hat{\sigma}_x \nonumber\\
&-
 \Delta(\bm{k}) \hat{\sigma}_y\, \hat{\tau}_y. \label{eq:BdG_continuous}
\end{align}
which enables us to derive the analytical expression of 
the quantized value of the conductance minimum.
We first apply a unitary trasformation to $H_{\mathrm{BdG}}$ to diagonalize 
the normal state Hamiltonian.
For $x$-type SOI with Eq.~\eqref{eq:subeq1}, we find
\begin{align}
&H_{\text{BdG}}(\bm{k}) = \check{U}\nonumber\\
&\times
\left[\begin{array}{cccc}
 \xi_{\bm{k}} + \lambda k_x & 0 & 0 &  \Delta(\bm{k})   \\
 0 & \xi_{\bm{k}} - \lambda k_x  & - \Delta(\bm{k}) & 0   \\
 0 &  - \Delta(\bm{k}) &  -\xi_{\bm{k}} + \lambda k_x & 0 \\
  \Delta(\bm{k}) &0& 0 &  -\xi_{\bm{k}} - \lambda k_x 
\end{array} \right] \nonumber \\
&\times\check{U}^\dagger, \\
&\check{U}= \frac{1}{\sqrt{2}} \left( 1 + i \hat{\sigma}_x\, \hat{\tau}_z\right).
\end{align}
The Hamiltonian is separated into the two $2\times 2$ Hamiltonian in this representation.
The $x$-type SOI divides the Fermi surface into two: 
one moves to $+k_x$ direction and the other moves to $-k_x$ direction.
The Fermi surface derived from the dispersion $\xi_{\bm{k}} + \lambda k_x$ ($\xi_{\bm{k}} - \lambda k_x$) 
is illustrated as an open circle labeled by the left (right) Fermi surface in Fig.~\ref{fig:fig5}(a) and (d), where a black dot 
indicates the $\Gamma$ point in the Brillouin zone (i.e., $\bm{k}=0$). 
The pair potentials for $d_{x^2-y^2}$-wave symmetry are shown on the two Fermi 
surfaces in Fig.~\ref{fig:fig5}(a).
In Fig.~\ref{fig:fig5}(a), Eq.~\eqref{eq:ABS} is satisfied at the shaded domains 
between the two dotted lines, where $k_y$ in such domains satisfies Eq.~\eqref{eq:nontrivial}.
The one-dimensional winding number at fixed $k_y$ is defined by \cite{sato:prb2011}
\begin{align}
\mathcal{W}(k_y) \equiv&
-\frac{1}{4\pi i} \int_{-\infty}^{\infty}dk_x \, I(k_y),\\
I(k_y)=& \mathrm{Tr}
\left[\hat{\tau}_x \, H_{\mathrm{BdG}}^{-1}\, \partial_{k_x}\, H_{\mathrm{BdG}} \right],\\
& \left\{ H_{\mathrm{BdG}}, \hat{\tau}_x \right\}_+=0. \label{eq:chiral_sym}
\end{align}
The one-dimensional winding number is calculated as
\begin{align}
\mathcal{W}(k_y) = \mathcal{W}(-k_y) = 2,
\end{align}
for all $k_y$ in the shaded domains in Fig.~\ref{fig:fig5}(a).  
As a result,  
\begin{align}
\mathcal{N}_{\mathrm{ZES}} = \sum_{k_y} \mathcal{W}(k_y),
\end{align}
remains a finite value.
This index represents the number of zero-energy states which form 
the resonant transmission channels in a DN.~\cite{ikegaya:prb2016}
The conductance for large enough $R_{\mathrm{N}}$ is quantized at
\begin{align}
G_{\mathrm{NS}} =  G_0 \times |\mathcal{N}_{\mathrm{ZES}}|. \label{eq:gns_min}
\end{align}
Thus $|\mathcal{N}_{\mathrm{ZES}}|$ increases by 4 with increasing $\lambda$ as
shown in Fig.~\ref{fig:fig3}. 
The index is approximately calculated as
\begin{align}
\mathcal{N}_{\mathrm{ZES}} \approx \left[ 2 N_c \, \frac{\lambda k_F}{ \mu}  \right]_{\mathrm{G}},
\label{eq:index1}
\end{align} 
 where $[ \cdots ]_\mathrm{G}$ means the integer part of the argument, $N_c$ is the number of 
 propagatting channels on the Fermi surface per spin.
 The expression is valid for $\lambda k_F \ll \mu$.
Details of the derivation are supplied in Appendix~\ref{sec:nzes}.
Thus the quantized value of the conductance increases monotonically with increasing $\lambda$.
In Fig.~\ref{fig:fig3}, the conductance jumps discontinuously because the number of propagating 
channels are limited for $W=25$ in the numerical simulation. 
Thus, $G_{\mathrm{NS}}$ will become a smoother function of $\lambda$ 
for a wider junction with $N_c \gg 1$.
In Fig.~\ref{fig:fig5}(d), we also illustrate 
 the pair potential for an $s$-wave SC with the $x$-type SOI. 
 Although the SOI splits the Fermi surface into two and modifies the amplitudes of the pair 
 potentials, it does not change the sign of the pair potentials. 
 Therefore, the APE is absent in $s$-wave junctions.
 The SOIs generate spin-triplet odd-parity Cooper pairs 
as a subdominant pairing correlation from the spin-singlet pair potentials.
This happens in both a $d$-wave SC and an $s$-wave SC. 
The induced spin-triplet Cooper pairs are expected to modify the properties of a 
spin-singlet SC to those in spin-triplet SC. 
The APE is a spin-triplet specific phenomenon.
 However, the appearance of the spin-triplet pairing correlation 
 is only a necessary condition for the APE. 
 Our results show that the APE requires the sign change in the pair potential.

 Thirdly, we discuss the effects of $y$-type SOI 
 with Eq.~\eqref{eq:subeq2} and those 
of Rashba SOI with Eq.~\eqref{eq:subeq3} on the pair potentials for $d_{x^2-y^2}$-wave symmetry.
The $y$-type SOI shift the Fermi surface $+ k_y$ ($-k_y$) direction and form the upper (lower) 
Fermi surface as shown in Fig.~\ref{fig:fig5}(b).
The condition Eq.~\eqref{eq:ABS} is not satisfied because the pair potentials 
are always even functions of $k_x$.
In the presence of Rashba SOI, 
the Fermi surfaces derived from the dispersion $\xi_{\bm{k}} + \lambda |\bm{k}|$ ($\xi_{\bm{k}} - \lambda |\bm{k}|$) 
forms the outer (inner) Fermi surface as illustrated in Fig.~\ref{fig:fig5}(c).
The condition Eq.~\eqref{eq:ABS} is not satisfied for the Rashba SOI because the pair potentials 
are always even functions of $k_x$.
Therefore, the APE is absent in these cases.

At the end of this section, 
we briefly discuss the effects of the misorientation of $d_{x^2-y^2}$-wave pair potential on the APE.
The pair potential in Eq.~\eqref{eq:delta_k} is expressed as $\Delta\, \cos 2\theta$ with 
$k_x=k_F \cos\theta$ and $k_y=k_F\sin\theta$.
When the pair potential is oriented as  $\Delta\, \cos 2(\theta-\beta)$ with $0 \leq \beta < \pi/4$,
the index in Eq.~\eqref{eq:index1} is calculated as
\begin{align}
\mathcal{N}_{\mathrm{ZES}} \approx \left[ 2 N_c \, \frac{\lambda k_F}{ \mu} 
\cos 2\beta \right]_{\mathrm{G}}.
\end{align} 
Therefore, the APE is robust to small misorientations of the pair potential.

\section{discussion}
\label{sec:discussion}
Finally, we discuss a relation between the conclusions of this paper and 
the experimental results in a T-shaped CoSi$_2$/TiO$_2$ junction on a Si substrate~\cite{lin:sciadv2021}.
In the experiment, a clear zero bias peak is observed in the conductance spectra, 
which means that a spin-triplet odd-parity (such as $p$-wave and $f$-wave) Cooper pair is 
definitely present in the thin film of CoSi$_2$.
The spin-triplet odd-parity pairing correlation makes the SC be topologically nontrivial 
which accommodates zero-energy quasiparticles at its surface. 
Majorana zero modes are a special case of such zero-energy states.  
The penetration of the zero-energy states into a dirty normal metal causes the APE.
 The zero-bias conductance peak in a T-shaped junction
reflects the enhancement of the density of states at zero energy due to the penetration.\cite{asano:prl2007tshape}
The index $\mathcal{N}_{\mathrm{ZES}}$ represents the 
degree of the degeneracy of such zero energy states in a dirty normal metal. 
We note, however, that spin-singlet $s$-wave superconductivity has been well established in bulk CoSi$_2$.
Therefore, theories that explain why spin-triplet Cooper pairs present in CoSi$_2$ are necessary.
The experiment also reported the strong SOIs at the interface between CoSi$_2$ and Si 
substrate.\cite{chiu:cjp2024,chiu:nanoscale2023}

To explane the experimental results, we are are compelled to consider symmetries other than $s$-wave symmetries 
for the pair potentials in CoSi$_2$ thin films on Si substrates, as listed below.
\begin{enumerate}
\item[(i)] The pair potential belongs to spin-triplet odd-parity class. 
\item[(ii)] Two pair potentials coexist. One belongs to spin-triplet odd-parity class.
 The other belongs to spin-singlet even-parity class. 
 \item[(iii)] Although the pair potential belongs to spin-singlet, SOIs generate 
spin-triplet Cooper pairs.
\end{enumerate}
If (i) is correct, our previous theory~\cite{asano:prl2007tshape} explains 
well the zero-bias peak in the experiment.
The authors of Refs.~\cite{mishra:prb2021, kirchner:prb2023} consider (ii)
and assume the effective attractive interactions at a $p$-wave channel under the strong SOIs.
The zero-bias peak in a T-shaped junction is explained only when the amplitude of the
spin-triplet pair potential is larger than that of the spin-singlet pair potential.
However, both (i) and (ii) require the spin-triplet $p$-wave pair potential.
At present, we have not known any $p$-wave attractive interactions between two electrons 
in CoSi$_2$. 
In this paper, we considered a different scenario of (iii) where the SC has only a 
spin-singlet pair potential.
We have shown in this paper that a $s$-wave SC does not indicate the APE with any types of SOIs. 
The spin-singlet $d$-wave pair potential with a specific SOI is necessary to explain the experiment.
The realization of spin-singlet $d$-wave order parameter might be easier than that 
of spin-triplet $p$-wave one.
However, we are not sure what mediates $d$-wave attractive interactions and 
if $x$-type SOI is realized in CoSi$_2$.
Therefore, the puzzle has not been solved yet.

\section{conclusion}
\label{sec:conclusion}
We theoretically studied the effects of the spin-orbit interaction (SOI) 
in a spin-singlet superconductor on the low-energy transport properties 
in a dirty normal-metal/superconductor junction as shown in Fig.~1. 
The differential conductance is calculated based on the Blonder-Tinkham-Klapwijk formula and 
the transport coefficients are calculated numerically by using the recursive Green's function method.
We consider two types of pair potentials such as $s$- and $d$-wave symmetry, and 
three types of SOI such as $x$-type, $y$-type, and Rashba type.
Our results demonstrate that a $d$-wave SC with $x$-type SOI exhibits the anomalous proximity effect (APE), 
whereas a $d$-wave SC with $y$-type SOI and that with Rashba SOI do not indicate the APE.
The numerical results also show that an $s$-wave SC with any types of SOI does not show 
the APE. 
We explain the numerical results by analyzing how SOIs change the sign of the pair potentials
on the Fermi surface.
%
Our findings provide an experimental setup for realizing an artificial spin-triplet SC. 

\begin{acknowledgments}
The authors are grateful to Y.~Tanaka, S.~Kirchner, and S.~Kobayashi for useful discussions.
J.~L. is supported by the Asahi Glass Foundation (No. S20K01).
S.~I. is supported by the Grant-in-Aid for Early-Career Scientists (JSPS KAKENHI Grant No. JP24K17010).
\end{acknowledgments}

\appendix

\section{Resistance in normal state}
 \label{sec:rn}
\begin{figure}[tttt]
\begin{center}
    \includegraphics[width=0.4\textwidth]{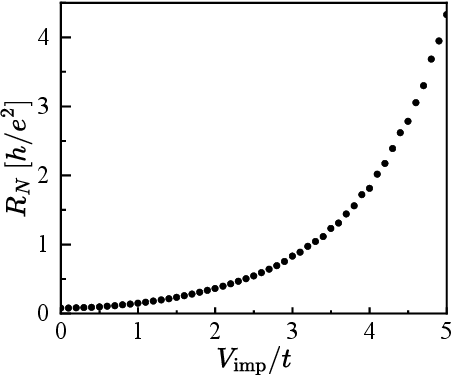}
    \caption{
The conductance is plotted as a function of the normal state resistance
in a normal metal $R_{\mathrm{N}}$ in Figs.~\ref{fig:fig2} and \ref{fig:fig4}. 
 }
    \label{fig:fig6}
\end{center}
\end{figure}

The conductance is plotted as a function of the normal state resistance
in a normal metal $R_{\mathrm{N}}$ in Figs.~\ref{fig:fig2} and \ref{fig:fig4}. 
Here, we explain how to obtain $R_{\mathrm{N}}$ in the numerical simulation.
We calculate the normal conductance $G_{\mathrm{N}}$ in Eq.~\eqref{eq:gn} 
for a disordered normal metal for $\mu = 2t$, $L = 50$, and $W = 25$. 
After repeating the calculation at a fixed $V_{\mathrm{imp}}$ over a number of 
different samples with different random potential configurations,  
$R_{\mathrm{N}}$ is defined by the inverse of the ensemble average of $G_{\mathrm{N}}$.
 In Fig.~\ref{fig:fig6}, we plot $R_{\mathrm{N}}$ as a function of $V_{\mathrm{imp}}$.
$R_{\mathrm{N}}$ increases with increasing $V_{\mathrm{imp}}$.
In in Figs.~\ref{fig:fig2} and \ref{fig:fig4}, we calculate the ensemble average of 
$G_{\mathrm{NS}}$ as a function of $V_{\mathrm{imp}}$ and plot the results 
as a function of $R_{\mathrm{N}}$.

\section{Atiyah-Singer Index}
 \label{sec:nzes}
We briefly summarize the relation between the quantized value of the zero-bias conductance and 
an index $\mathcal{N}_{\text{ZES}}$. Let us begin with a BdG Hamiltonian 
for a spin-singlet SC with the $x$-type SOI in Eq.~\eqref{eq:BdG_continuous}.
The wave number on the Fermi surface is determined by
\begin{align}
\xi_k + s \, \lambda\,  k_x = 0, \quad  s = \pm 1,
\end{align}
where $s=1(-1)$ corresponds to the Fermi surface shifted to left (right) in Fig.~\ref{fig:fig5}(a).
The wave numbers in the two directions satisfy
\begin{align}
(k_x + s \tilde{\lambda} k_F)^2 + k^2_y = (1 + &\tilde{\lambda}^2) k^2_F , \\
\tilde{\lambda} \equiv \frac{\lambda k_F}{2 \mu} \ll 1.&
\end{align}
The $x$-type SOI shifts the center of the Fermi surface to $\pm \tilde{\lambda} k_F$ on the $k_x$ axis.
The pair potential of a $d_{x^2-y^2}$-wave SC on the Fermi surface has nodes at $k_x = \pm k_y$.
As a result, Eq.~\eqref{eq:ABS} is satisfied at the channels 
\begin{align}
 \frac{k_F}{2} (\sqrt{\tilde{\lambda}^2 +2} - \tilde{\lambda}) \le |k_y | \leq 
 \;\frac{k_F}{2} (\sqrt{\tilde{\lambda}^2 +2} + \tilde{\lambda}),
 \label{eq:nontrivial}
 \end{align}
for both the left and the right Fermi surfaces.
The schematic figure of the pair potentials are shown in Fig.~\ref{fig:fig5}(a).
The pair potential indicated by shaded area in Fig.~\ref{fig:fig5}(a) satisfies 
Eq.~\eqref{eq:ABS}, indicating the appearance of surface Andreev bound state at each propagating channel.
The number of such ZESs for the two spin sectors is estimated as 
\begin{align}
N= \left[4 N_c  \tilde{\lambda}\right]_\mathrm{G},
\end{align} 
where $N_c = W k_F / \pi$ is the number of propagating channels for each spin sector, $W$ is the width of the SC in the $y$ direction, and $[\ldots]_G$ means the Gauss symbol providing the integer part of a number. 
As a result, the number of Andreev bound states at zero-energy
is $N$ at a clean surface of a $d_{x^2-y^2}$-wave SC with the $x$-type SOI.
In other words, $N$ represents the degree of the degeneracy of zero-energy states at a surface of a SC.
Such a high degeneracy is a result of translational symmetry in the $y$ direction of a clean SC.
In a clean normal-metal / SC junction, the zero-energy states penetrate into the clean normal metal 
and form the perfect transmission channels. 

\begin{figure}[tttt]
\begin{center}
    \includegraphics[width=0.47\textwidth]{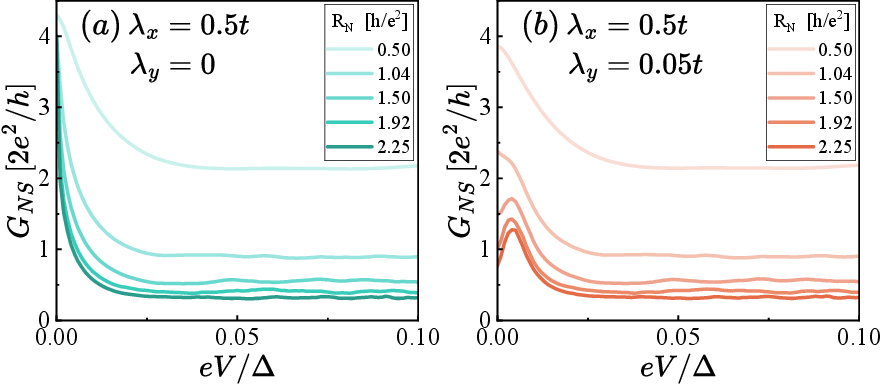}
    \caption{The conductance of a $d_{x^2-y^2}$-wave junction is plotted
 as a function of the bias voltage for several $R_N$.
 The results for the $x$-type at $\lambda_x=0.5t$ are shown in (a). 
 In (b), we choose $\lambda_x=0.5t$ and $\lambda_y=0.05t$. 
 }
    \label{fig:fig7}
\end{center}
\end{figure}

To discuss effects of random potentials in a normal metal attached to a SC, 
the analysis of chiral property of zero-energy states is necessary.\cite{ikegaya:prb2015,ikegaya:prb2016}  
The BdG Hamiltonian in Eq.~\eqref{eq:BdG_continuous} preserves chiral symmetry
in Eq.~\eqref{eq:chiral_sym}.
Since $\hat{\tau}_x^2=1$, the eigenvalues of $\hat{\tau}_x$ are either
1 (positive chirality) or -1 (negative chirality). 
It is known that a zero-energy state of $H_{\text{BdG}}$ is the eigenstate of $\hat{\tau}_x$.
Therefore, such a zero-energy state has either the positive chirality or the negative chirality.
The wave functions of the zero-energy states are calculated as
\begin{align}
\psi_{1} = 
\begin{pmatrix}
i \\
1 \\
-i \\
-1
\end{pmatrix}
A e^{ik_y y} f(x),
\;
\psi_{2} = 
\begin{pmatrix}
1 \\
i \\
-1 \\
-i
\end{pmatrix}
A e^{ik_y y} f(x)
\end{align}
where $A$ is a normalization constant and $f(x)$ is a function localizing at a surface of a SC.
It is easy to confirm that all of the zero-energy states 
belong to the negative chirality, as it satisfies 
$\hat{\tau}_x \, \psi_{j} = - \psi_{j}$ for $j=1$ and 2. 
The Atiyah-Singer index is defined by
\begin{align}
\mathcal{N}_{\text{ZES}}= N_+ - N_-,
\end{align} 
where $N_+$ ($N_-$) is the number of zero-energy states belonging to positive (negative) chirality. 
Therefore, the index is calculated as
\begin{align}
|\mathcal{N}_{\text{ZES}}|= \left[  4 N_c  \tilde{\lambda} \right]_G . \label{eq:nzse5}
\end{align} 
The index $\mathcal{N}_{\text{ZES}}$ is an invariant in the presence of chiral symmetry of the Hamiltonian.
Here we calculate the index in a clean SC by assuming the translational symmetry in the $y$ direction.
The index remains unchanged even when the random impurity potentials
\begin{align}
H_{\mathrm{imp}}=V(\bm{r}) \, \tau_z,
\end{align}
enters the Hamiltonian in Eq.~\eqref{eq:BdG_continuous}.
This is because the random potential preserves chiral symmetry.
In physics, $|\mathcal{N}_{\text{ZES}}|$ represents the number of zero-energy states 
that penetrate into a dirty normal metal while retaining their high degeneracy 
and form the perfect transmission channels. 
The electric current through such perfect transmission channels is independent of $R_{\mathrm{N}}$, 
whereas the electric current through usual transmission channels decreases with increasing $R_{\mathrm{N}}$.
As a result, the minimum value of the conductance at zero bias is described 
by Eq.~\eqref{eq:gns_min}~\cite{ikegaya:prb2015, ikegaya:prb2016}.

\section{$x$-type SOI to Rashba SOI}
\label{sec:ytype}
We briefly discuss the effects of the $y$-type SOI introduced 
in addition to the $x$-type SOI. 
The BdG Hamiltonian in continuous space is 
given in Eq.~\eqref{eq:BdG_continuous} with $\lambda_z \neq 0$ and $\lambda_y \neq 0$.
It is easy to confirm that the BdG Hamiltonian does not preserve 
chiral symmetry in Eq.~\eqref{eq:chiral_sym}.
Therefore, the index $\mathcal{N}_{\mathrm{ZES}}$ can no longer be defined
and the conductance deviates from its quantized value.
In Fig.~\ref{fig:fig7}, the conductance is plotted as a function of the bias 
voltage for several choices of $R_{\mathrm{N}}$. 
We choose $(\lambda_x, \lambda_y)=(0.5 t, 0)$ in (a) and 
$(\lambda_x, \lambda_y)=(0.5 t, 0.05t)$ in (b). 
The minimum value of the zero-bias conductance is quantized 
as Eq.~\eqref{eq:gns_min} in Fig.~\ref{fig:fig7}(a) 
irrespective of $R_{\mathrm{N}}$. 
When we add the $y$-type SOI, 
the zero-bias conductance decreases gradually with 
increasing $R_{\mathrm{N}}$ as shown in Fig.~\ref{fig:fig7}(b).
When $\lambda_y$ is increased to $\lambda_x$, the results 
for Rashba SOI in Fig.~\ref{fig:fig2} do not exhibit any indication of the APE.
Thus, the APE is fragile under perturbations that break chiral symmetry.

\bibliographystyle{apsrev4-2}
%

\end{document}